\documentclass[usegraphicx,useAMS,usenatbib,fleqn]{mn2e}
\include{latex_macros}
\usepackage{float}

\usepackage{times}
\usepackage{amsmath}
\usepackage{amssymb}
\usepackage{amsfonts}
\usepackage{mathrsfs}
\title {A search for radius inflation among active M-dwarfs in Praesepe}
\author[R. J. Jackson, R. D. Jeffries, C. P. Deliyannis, Q. Sun, S. T. Douglas]
  {R. J.~Jackson$^1$, R. D.~Jeffries$^1$, Constantine
    P.~Deliyannis$^2$, Qinghui Sun$^2$ \newauthor and Stephanie.~T.~Douglas$^{3,4}$\\
   $^1$ Astrophysics Group, Keele University, Keele, 
      Staffordshire ST5 5BG\\
     $^2$Department of Astronomy, Indiana University.
727 E 3rd Street,
Bloomington, IN 47405-7105, USA\\
$^3$ Harvard-Smithsonian Center for Astrophysics, 60 Garden Street, Mailstop 15, Cambridge, MA, USA\\
$^4$ NSF Astronomy \& Astrophysics Postdoctoral Fellow\\
}

\date{Accepted for publication}

\pagerange{\pageref{firstpage}--\pageref{lastpage}} \pubyear{2018}

\def\LaTeX{L\kern-.36em\raise.3ex\hbox{a}\kern-.15em
    T\kern-.1667em\lower.7ex\hbox{E}\kern-.125emX}

\begin{document}
\label{firstpage}
\maketitle

\begin{abstract}
Rotation periods from Kepler K2 are combined with projected
rotation velocities from the WIYN 3.5-m telescope, to determine 
projected radii for fast-rotating, low-mass ($0.15 \leq M/M_{\odot} \leq
0.6$) members of the Praesepe cluster. A maximum likelihood analysis
that accounts for observational uncertainties, binarity
and censored data, yields marginal evidence
for radius inflation -- the average radius of these stars is $6\pm4$
per cent larger at a given luminosity than predicted by commonly-used
evolutionary models. This over-radius is smaller (at 2-sigma
confidence) than was found for similar stars in the younger Pleiades 
using a similar analysis; any decline appears
due to changes occurring in higher mass ($>0.25 M_{\odot}$)
stars. Models incorporating magnetic inhibition of
convection predict an over-radius, but do not reproduce this mass
dependence unless super-equipartition surface magnetic fields are
present at lower masses. Models incorporating flux-blocking by
starspots can explain the mass dependence but there is no evidence that
spot coverage diminishes between the Pleiades and Praesepe
samples to accompany the decline in over-radius. The fastest rotating
stars in both Praesepe and the Pleiades are significantly smaller than the
slowest rotators for which a projected radius can be measured.  This
may be a selection effect caused by more efficient angular momentum
loss in larger stars leading to their progressive exclusion from the
analysed samples. Our analyses assume random spin-axis orientations;
any alignment in Praesepe, as suggested by Kovacs (2018), is strongly
disfavoured by the broad distribution of projected radii.
\end{abstract}


\section{Introduction}

There are significant differences between the predictions of stellar
models and the precisely measured masses and radii of
main-sequence K- and M-dwarfs in
eclipsing binary systems. For a given mass, the absolute radii
of some stars
with $0.2< M/M_{\odot}<0.8$ are 10--20 per cent larger than predicted
and hence, for a given luminosity, the effective temperature, $T_{\rm
  eff}$ may be under-estimated by 5--10 per cent (e.g. Lopez-Morales \&
Ribas 2005; Morales et al. 2009; Torres 2013). Since the
interferometrically determined radii of nearby, slowly rotating
low-mass stars are in better agreement with ``standard''
evolutionary models, it has been hypothesised that the enlarged radii of
fast-rotating binary components, are due to dynamo-generated
magnetic activity; either through inhibiting convection
(e.g. Mullan \& MacDonald 2001; Feiden \& Chaboyer 2014) or by 
blocking outward flux with dark starspots (e.g. Spruit \& Weiss 1986;
MacDonald \& Mullan 2013; Jackson \& Jeffries 2014a).

There is growing indirect evidence that this phenomenon also occurs
in fast-rotating young pre main sequence (PMS)
and zero-age main sequence (ZAMS) stars and may play a role in the
explanation of several astrophysical problems affecting stars
with deep convective envelopes, rapid rotation and high levels of
magnetic activity. These include the anomalously red colours of
PMS/ZAMS stars; the rotation-dependent scatter in lithium depletion
seen in young stars of otherwise similar mass and age; and
disagreements between measured mass, radius and position in the
Hertzsprung-Russell (HR) diagram for several PMS/ZAMS eclipsing binaries (see
for example Somers \& Pinsonneault 2014; Jackson \& Jeffries 2014a;
Covey et al. 2016; Feiden 2016; Kraus et al. 2016, 2017; Jeffries et
al. 2017; Somers \& Stassun 2017; Bouvier et al. 2018).

Direct determination of the radii of young stars by interferometric techniques
is hampered by their distance. An indirect approach measures the
projected radii of stars by combining the rotation period ($P$ in days)
with the projected equatorial velocity ($v \sin i$ in
km\,s$^{-1}$):
\begin{equation}
\frac{R\sin i}{R_{\odot}} = 0.0198\, P\, v \sin i\, 
\end{equation} 
(e.g. Rhode, Herbst \& Mathieu 2001; Jeffries 2007).
If the spin axis orientation of the stars is random (e.g. Jackson \& Jeffries
2010) and observational biases are understood, then a set of $R \sin i$ estimates can be modelled to
determine the true average radius for a group of stars. This technique has been
used to claim the detection of inflated radii in the young K- and
M-type stars of several young clusters, notably the Pleiades at an age
of $\simeq 120$ Myr (Jackson, Jeffries \& Maxted 2009; Hartman et
al. 2010; Jackson \& Jeffries 2014a; Jackson et al. 2016; Lanzafame et
al. 2017).

\begin{figure*}
	\centering
	\begin{minipage}[t]{0.93\textwidth}
	\centering
	\includegraphics[width = 180mm]{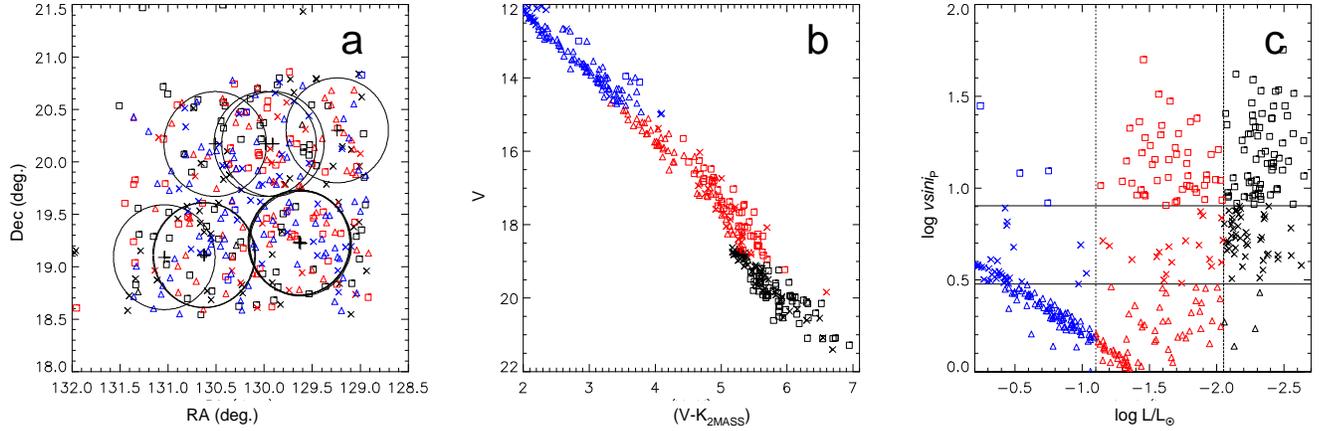}
	\end{minipage}
	\caption{Target selection: Panel (a) shows the spatial
          distribution of targets. Circles show the field of view of the eleven
          observed WIYN Hydra configurations (see Table 1).  The
          different symbols correspond to divisions in
          estimated luminosity shown in panels (b) and (c), and the different
          symbol shapes correspond to the predicted observed $(v\sin
          i)_p$ (see section 2.1) shown in panel (c) -- triangles
          indicate $(v\sin i)_p < 3$km\,s$^{-1}$, crosses have $3<
          (v\sin i)_p <8$ km\,s$^{-1}$ and squares $(v\sin i)_p \geq 8$
          km\,s$^{-1}$). Panel (b) shows $V$ versus $V-K_{\rm
            2MASS}$ colour magnitude diagram for the same data set
          using the same symbols and colour coding. Panel (c) shows $(v
          \sin i)_p$ as a function of estimated target luminosity.}
	\label{fig1}
\end{figure*}

In Jackson et al. (2018, hereafter Paper I), we obtained the largest
ever set of homogeneously determined values of $v \sin i$ and $P$ for
low-mass stars in the Pleiades to investigate radius inflation as a
function of mass using Eqn.~1. It was
established that stars with $0.1<M/M_{\odot}<0.8$ are $14\pm 2$ per
cent larger than predicted by standard evolutionary models (e.g. Dotter
et al. 2008; Baraffe et al. 2015) at a given luminosity, with no
evidence for a strong mass-dependence. The lack of mass-dependence is
incompatible with published evolutionary models that incorporate the
inhibition of convection by magnetic fields
(e.g. Feiden, Jones \& Chaboyer 2015),
but might be consistent with a mass-dependent starspot coverage or 
some combination of the two effects.

A critical diagnostic would be to see whether radius inflation is only
associated with the rapid rotation of these low-mass stars, or whether
their age and hence evolutionary stage and interior structure are
important. A very similar analysis of a heterogeneous (but presumably much
older than the Pleiades) sample of fast-rotating field M-dwarfs also
found inflation levels of $14 \pm 3$ per cent compared with
non-magnetic evolutionary models, suggesting that age
itself may not be a vital parameter (Kesseli et al. 2018).  In this
paper, we extend the work of Paper I to a homogeneous population of
coeval, fast-rotating low-mass stars in the older ($\simeq 650$ Myr)
Praesepe cluster, with measured rotation periods from the Kepler K2
mission and new measurements of $v \sin i$ using the WIYN-3.5m
telescope and Hydra spectrograph. Whilst M-dwarfs in the Pleiades are
still in the PMS phase (or have just reached the ZAMS), stars of
equivalent mass in Praesepe should be firmly established on the
hydrogen-burning main sequence and those with mass $M\geq
0.35M_{\odot}$ should have radiative cores (Baraffe et al. 2015).

\section{Spectroscopic observations}

\subsection{Target selection}
Candidate targets were selected from a list of high probability
Praesepe members (from Kraus \& Hillenbrand 2007) with rotation periods
reported by Douglas et al. (2017). The large majority of these periods
(92 per cent) are based on K2 light curves from campaign 5 (Howell et
al. 2014) and the completeness of the rotation period data is 86
per cent. Stars with rotation periods were matched with the 2MASS
catalogue (Skrutskie el al. 2006) to define the target names,
co-ordinates and $K_{\rm 2MASS}$, and with the
Gaia catalogue (DR1, Gaia collaboration 2016)
for $G$ magnitudes. Figure~1a shows
the spatial distribution of potential targets. Targets for our
fibre-spectroscopy were selected from a 10 square degree area
with the highest target density, with a faint limit
of $K_{\rm 2MASS}<14.5$.

The luminosity of each target was estimated from its $K_{\rm CIT}$
magnitude (hereafter referred to as $K$)
assuming a conversion of $K = K_{\rm 2MASS}+0.024$ (Carpenter
2001), zero reddening (Cummings et al. 2017), a distance modulus of
6.35$\pm$0.04 (Gaia Collaboration et al. 2018) , and taking bolometric
corrections as a function of  $(V-K)_0$ from a 625\,Myr Baraffe et al. (2015;
hereafter BHAC15) isochrone. $V$ magnitudes were only available for
54 per cent of the
stars. This subset was used to define a second-order polynomial
relationship between $V-K$ and $G-K$ 
(shown in Fig.~2) that allows the assignment of $V-K$, bolometric
corrections and luminosities to the other stars, with an rms
uncertainty of $\pm 0.17$ mag in $V-K$, corresponding to $\pm 0.07$ dex
in log luminosity. The colour-magnitude diagram for potential targets
is shown in Fig.1b.

Targets were prioritised according to a predicted
projected equatorial velocity, $(v \sin i)_p =50\, (\pi /4)\,R/P$ in
km\,s$^{-1}$, where $R$ is the
stellar radius in solar units estimated from the 625\,Myr BHAC15
isochrone and $\pi/4$ is a simple average value for the
(unknown) $\sin i$ if
the spin axes are randomly oriented.
Targets with $(v \sin i)_p >8$\,km\,s$^{-1}$ were given the
highest weighting for target selection, since these were likely to
yield a resolvable $v\sin i$ (see Fig~1c). The BHAC15 models are calculated
at a solar metallicity, but Praesepe has a super-solar metallicity
([Fe/H]=0.156$\pm$0.004 -- Cummings et al. 2017). The effect of
metallicity on the estimated over-radius is considered in section 4.

\begin{table*}
\caption{Hydra Configurations observed in Praesepe. The positions
  are those of the field centres.}
\begin{tabular}{ccccccccccc} \hline
 Config. & File & Hydra  & Date & UT of     &RA      & Dec   & Number~~& Total Exp.  & Fibres on & Fibres on \\
 number & number& fibres&    & exposure \#1 & \multicolumn{2}{c}{(J2000)} & exposures & time (s)& targets  & sky \\\hline
1  &        4058 & blue & 2017-01-03 & 10:16:59.0 & +08:39:37.70       & +20:09:58.260  &  4  &  14400  &  46  &  25\\						
2  &        6070 & blue& 2017-01-05 & 10:41:43.0 & +08:42:32.20       & +19:06:28.27  &  3  &  9285  &  39  &  27\\								
3  &       13036 & blue& 2017-01-18 & 08:54:37.0 & +08:38:00.14      & +19:13:29.25  &  2  &  4028  &  39  &  22\\							
4  &       14030 & blue& 2017-01-19 & 07:37:29.0 & +08:38:00.12      & +19:13:32.84  &  2  &  6436  &  42  &  27\\							
5  &       21084 & blue& 2017-02-02 & 07:47:06.0 & +08:42:32.65      & +19:06:29.63  &  2  &  7200  &  38  &  26\\							
6  &       21086 & blue& 2017-02-02 & 10:13:11.0 & +08:38:00.97      & +19:13:30.30  &  2  &  8100  &  45  &  26\\							
7  &       22081 & blue& 2017-02-03 & 07:48:29.0 & +08:42:36.24      & +20:10:01.95  &  5  &  17400  &  34  &  25\\							
8  &       36069 & red & 2017-02-21 & 02:47:19.0 & +08:39:37.41      & +20:09:57.23  &  4  &  14400  &  53  &  25\\							
9  &       36073 & red & 2017-02-21 & 07:10:48.0 & +08:38:00.12      & +19:13:30.54  &  4  &  14400  &  51  &  25\\							
10 &       37050 & red & 2017-02-22 & 02:46:01.0 & +08:36:31.67      & +20:17:58.31  &  4  &  14400  &  31  &  24\\							
11 &       37055 & red & 2017-02-22 & 07:26:14.0 & +08:44:27.98      & +19:05:08.42  &  4  &  13500  &  28  &  25\\\hline        
    \end{tabular}
  \label{observations}
\end{table*}

\begin{table*}
\caption{Properties of observed science targets in the
  Praesepe. Periods are from Douglas et al. (2017), $BC_K$, $\log L/L_{\odot}$, $M/M_{\odot}$  and $R/R_{\odot}$ and 
 a {\it predicted} $(v \sin i)_p$  are estimated using a BHAC15 model isochrone - see section 2.1. A sample of the  table is given here, the full table is made available electronically.}
\begin{tabular}{lccccccccccc} \hline
Target name		&RA	&Dec	&K$_{\rm 2MASS}$&(V-K)$_0$ &source  &Period &$BC_{\rm K}$	&$\log L/L_{\odot}$ & $M/M_{\odot}$
&$R/R_{\odot}$ &$(v \sin i)_p$\\
(2MASS)		&\multicolumn{2}{c}{(J2000)}		  & (mag)   	&(mag)*& (V-K)&(days)&(mag)&&& &(km\,s$^{-1}$)\\\hline
J08355651+2037070	&	08 35 56.510	&	+20 37 07.03	&	14.08	&	5.78	&	Gmag	&	1.22	&	2.82	&	-2.33	&	0.20	&	0.215	&	7.0	\\
J08360242+1904265	&	08 36 02.423	&	+19 04 26.52	&	14.25	&	5.52	&	Gmag	&	0.33	&	2.79	&	-2.38	&	0.19	&	0.206	&	24.8	\\
J08364501+2008459	&	08 36 45.016	&	+20 08 45.90	&	13.78	&	5.30	&	Gmag	&	2.23	&	2.77	&	-2.19	&	0.24	&	0.246	&	4.4	\\
J08364895+1918593	&	08 36 48.959	&	+19 18 59.35	&	11.95	&	4.63	&	Vmag	&	1.17	&	2.67	&	-1.42	&	0.51	&	0.471	&	16.0	\\
J08365162+1850193	&	08 36 51.627	&	+18 50 19.31	&	13.48	&	5.35	&	Gmag	&	2.06	&	2.77	&	-2.07	&	0.28	&	0.271	&	5.2	\\
J08393071+1856533	&	08 39 30.714	&	+18 56 53.40	&	13.84	&	5.40	&	Gmag	&	1.04	&	2.78	&	-2.22	&	0.23	&	0.239	&	9.1	\\
J08395128+2034499	&	08 39 51.288	&	+20 34 49.97	&	11.59	&	4.62	&	Vmag	&	1.84	&	2.67	&	-1.27	&	0.56	&	0.517	&	11.2	\\\hline
\multicolumn{10}{l}{* (V-K)$_0$ colour estimated from G-K$_{\rm 2MASS}$ colour where no V magnitude is available (see section 2.1).}\\    \end{tabular}
  \label{targets}
\end{table*}

\subsection{Observations}


Observations of the Praesepe targets were made in a similar manner to
that described in Paper I at the WIYN 3.5-m telescope 
using the WIYN Hydra multi-object spectrograph
(Bershady et al. 2008). 
Eleven configurations were observed in Praesepe over the period from
3 January 2017 to 22 February 2017. Field centres were chosen to maximise the
number of highest priority targets. The first 7 configurations were
observed using the ``blue" Hydra fibres giving a resolving power of 14000
(determined from the line width of the arc spectra). These observations
were made on the same nights and with the same telescope set up as
previously reported observations of Pleiades targets (Paper I). The
final 4 configurations were observed using
the "red" Hydra fibres giving an increased resolving power of
19000. Spectra were recorded over a $\sim 400$\AA\ interval, centred at
$\sim 7880$\AA. The FWHM of a resolution element was sampled by about 2.5
(binned) pixels. Details of the configuration and exposure times
are given in Table 1. Long exposures were split into multiples
of about 1 hour to aid cosmic ray rejection. A total of
446 spectra were measured for 230 unique targets in Praesepe, 283 using
blue Hydra fibres and 163 using the red fibres. Properties of the observed
targets are listed in Table 2.

\begin{figure}
	\centering
	\includegraphics[width = 75mm]{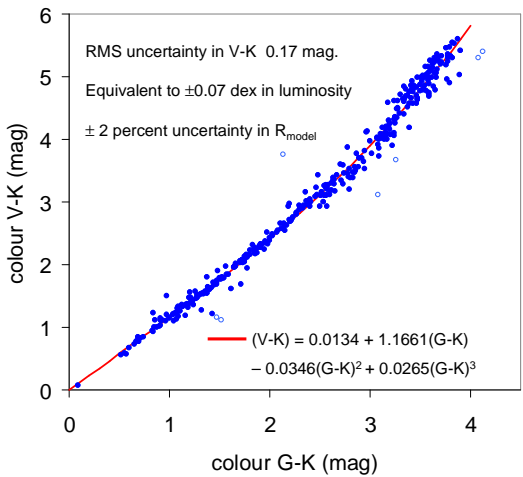}
	\caption{Colour colour plot for targets in Praesepe with reported values of $V$, $G$ and $K_{\rm 2MASS}$ magnitudes. 
	The red line shows the fitted cubic relation between the two colours. This relation is used to
	estimate the $V-K$ for Praesepe targets with no reported $V$ magnitude.}
	\label{fig2}
\end{figure}
\begin{figure}
	\centering
	\includegraphics[width = 75mm]{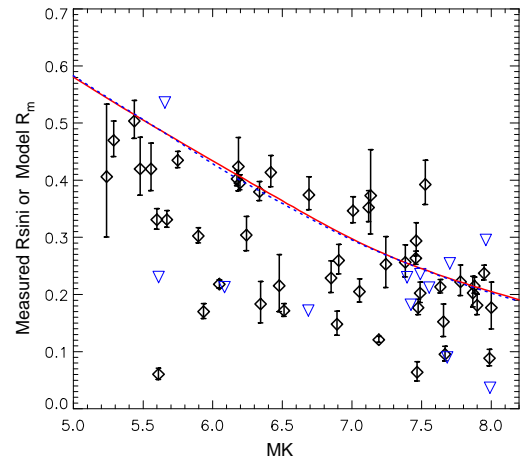}
	\caption{Projected radii, $R\sin i$ (in solar units), vs $M_K$ for low mass
          stars in Praesepe. Diamonds show targets with
          $R\sin i$ uncertainties $\le 30$ per
          cent. Triangles show upper limit values for targets with
          larger uncertainties. Solid and dashed lines are
          predictions from 625\,Myr BHAC15 and
          Dartmouth (Dotter et al. 2008) solar metallicity isochrones respectively.}
	\label{fig3}
\end{figure}

\subsection{Data reduction}
Many targets were faint, requiring optimal extraction
of their spectra to provide sufficient signal-to-noise ratio (SNR) for useful
analysis. Strong sky emission lines dominated the
fainter spectra. For these reasons we used purpose-built software for
data reduction, described in detail in Paper I, but summarised here. 
De-biased science data images
were normalised to daytime tungsten lamp flat field exposures. Spectra
were extracted from the normalised image using an optimal extraction
algorithm (Horne 1986). Daytime Th-Ar lamp exposures were
used to define polynomial relations between pixels in the
extracted spectra and wavelength, and a correction applied to the
target exposures based on the position of prominent sky emission lines
in the median blank sky spectra. The spectra were
rebinned to a common wavelength range and sky subtracted using
median sky spectra weighted according to the measured fibre
efficiencies. Spectra from repeat exposures, after heliocentric
velocity correction, were then stacked to produce final target spectra
covering a wavelength range 7681--8095\AA\ in 0.1\AA~steps.

\subsection{Radial velocity and projected equatorial velocity}

The estimation of radial velocity ($RV$) and projected equatorial velocity ($v\sin i$) 
was achieved by cross correlating target spectra
with template spectra from the UVES atlas (Bagnulo et al. 2003, see
Table~3) after truncation shortward of 7705\AA\ to exclude strong
telluric features. The adopted methods were similar to those described in
Paper I: $RV$ and $v \sin i$ were determined from the peak and FWHM of
a Gaussian profile fitted to the cross-correlation function (CCF). The
increase in FWHM with respect to that determined for a set of
slowly rotating stars of similar spectral type (FWHM$_0$) was calibrated
by artifically broadening the spectra of bright standard stars (see
Table~3). Independent calibration curves were determined for the red-
and blue-fibre setups.


The precision of $RV$, FWHM and $v\sin i$ measurements were calculated
according to equations (2)--(4) derived in Paper I, where the
measurement precision is defined as the product of a scaling factor
$S$ and a t-distribution with $\nu$ degrees of freedom.
The constants used in these expressions were determined empirically
from repeated observations of 174 targets in the Pleiades and Praesepe taken with blue fibres. There
were too few repeat observations made with red fibres to make an
independent estimate of the constants in the scaling formulae so the
same values were adopted. This is acceptable since the principal
difference between spectra measured using the red and the blue fibres
is the change in resolution and this is already accounted for by the
presence of FWHM and FWHM$_0$ in the scaling formulae.

Table 4 gives the weighted (by $S^{-2}$) mean $RV$ and $v\sin i$ values, and
their estimated uncertainties, for 230 independent targets in Praesepe.
The $RV$s are measured relative to the median $RV$; no attempt to provide
 an absolute RV calibration was
made.  The dispersion of the relative $RV$s for red and blue fibres
estimated from the median absolute dispersion (MAD) of the target
$RV$s, are $\sim$1.0\,km\,s$^{-1}$ and $\sim$ 1.5\,km\,s$^{-1}$
respectively (using the approximate relation $\sigma_{t}=$MAD/0.68 for
a t-distribution with $\nu=4$). These values represent the combined
effects of (a) intrinsic dispersion in the cluster, (b) measurement
uncertainties and (c) the effects of binarity and are therefore upper
limits to the intrinsic velocity dispersion of targets within the
cluster.

\begin{table}
\caption{Calibration standards used as cross-correlation function (CCF)
  templates to determine RVs (see section 2.4). Spectra of the
  $v\sin i$ standard stars were used to define calibration curves of
  $v\sin i$ versus the increase in FWHM relative to the value measured
  for slowly rotating stars, FWHM$_0$, for the blue and red fibres respectively.}

    \begin{tabular}{llllll} \hline
No. &M$_{\rm K}$  & CCF template         &$ v\sin i$ & \multicolumn{2}{c}{FWHM$_0$ (\,km\,s$^{-1}$)}\\
 & range    &      &  standards & blue & red\\\hline
1 & $>$5.5	  & HD~34055		      &	Gl\,133/Gl\,285 & 24.20 & 21.30\\
2 & 4.9 - 5.5 & HD~130328 	  &	Gl\,133/Gl\,285 & 24.80 & 22.35 \\
3 & 4.4 - 4.9 & HD~156274		    &	Gl\,184/Gl\,205 & 33.35 & 26.90 \\\hline
    \end{tabular}
  \label{standards}
\end{table}

\begin{table*}
\caption{Estimated values of relative $RV$, $v\sin i$ and $R\sin i$
  (from Eqn.~1). Absolute uncertainties in $RV$ and $v\sin i$ (defined
  according to Eqns.~2 and 4) are also given. $R\sin i$ values are shown where $(v\sin i)_p$$>$8\,km\,s$^{-1}$. Where the relative uncertainty in $v\sin i$ 
is $>30$ per cent, an upper limit is shown along with corresponding
upper limits in $R\sin i$. A sample of the Table is shown here, the full version is available electronically.}
\begin{tabular}{lrrrrrrrrrr} \hline
Target name &	$M_K$ &$\log L/L_{\odot}$&Period
&SNR&$RV_{\rm rel}$&$S_{RV}$&$v\sin i$&$S_{v\sin i}$&$R\sin i$\\
(as 2MASS)  &(mag)&      &(d)	&   &(km\,s$^{-1}$)&(km\,s$^{-1}$)&(km\,s$^{-1}$)&(km\,s$^{-1}$)&($R_{\odot}$) \\\hline		
J08355651+2037070	&	7.73	&	-2.33	&	1.22	&	22	&	-0.80	&	1.19	&	10.4	&	2.0	&	---	\\
J08360242+1904265	&	7.90	&	-2.38	&	0.33	&	10	&	0.40	&	3.65	&	27.7	&	2.4	&	0.181	\\
J08364501+2008459	&	7.43	&	-2.19	&	2.23	&	15	&	0.50	&	4.87	&	14.0	&	6.2	&	---	\\
J08364895+1918593	&	5.60	&	-1.42	&	1.17	&	140	&	1.33	&	0.51	&	14.3	&	0.7	&	0.331	\\
J08365162+1850193	&	7.13	&	-2.07	&	2.06	&	27	&	2.28	&	0.91	&	7.6	&	5.0	&	---	\\
J08393071+1856533	&	7.49	&	-2.22	&	1.04	&	26	&	0.60	&	2.03	&	9.0	&	4.3	&	$<$0.162	\\
J08395128+2034499	&	5.24	&	-1.27	&	1.84	&	69	&	0.06	&	0.58	&	11.2	&	2.9	&	0.407	\\
\hline
\end{tabular}
\label{vsini}
\end{table*}

\begin{figure*}
	\centering
	\includegraphics[width = 170mm]{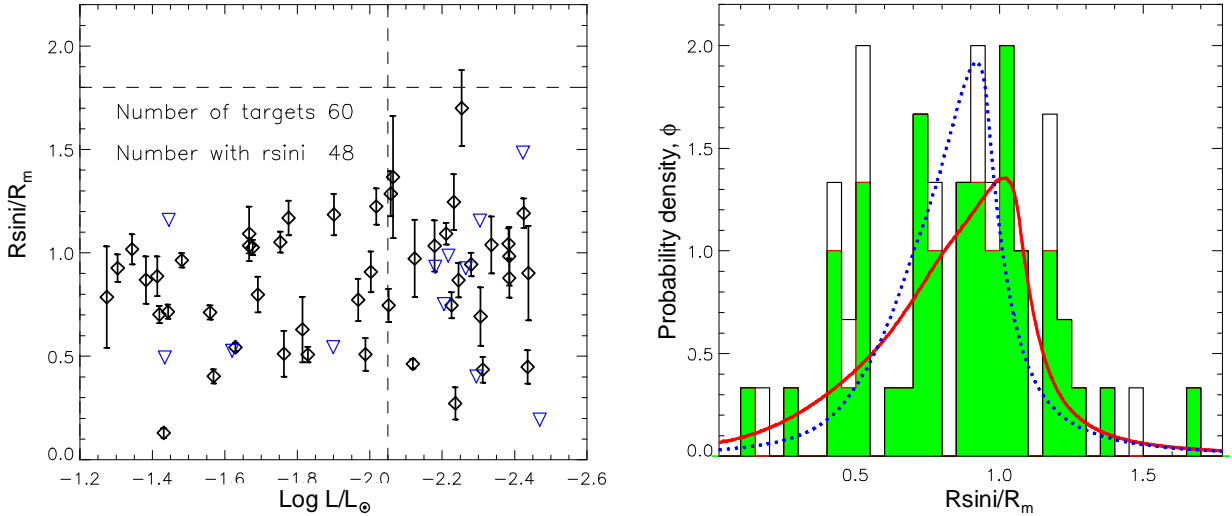}
	\caption{Normalised radii of  low mass stars in Praesepe. Plot (a) shows $r\sin i$ as a function of luminosity. Diamonds show $r\sin i$ for targets with a relative uncertainty $\le$30 per cent. Triangles show upper limits for targets        $(v\sin i)_p  >8$~km\,s$^{-1}$ with higher levels of  uncertainty. Plot (b) shows the number
          density of targets as a function of $r \sin i$. Targets with
          a relative uncertainty  $\le 30$ per cent are shown as a solid histogram. The
          open histogram including the stars with upper limits at their
          upper limit values. The dotted line shows a model
          distribution with no over-radius (using BHAC15 model radii and a 
          random distribution of stellar spin-axis orientation). The solid line shows the maximum likelihood model for
          the best-fitting over-radius of $\rho = 1.07 \pm 0.03$. }
\label{fig4}
\end{figure*}

\section{Comparison of measured radii with current evolutionary models}
Equation~1 is used with the measurements of $P$ and $v\sin i$ in Table 4 to estimate
$R\sin i$ for targets with $|RV_{\rm
  rel}|<10$\,km\,s$^{-1}$ and $(v \sin i)_p > 8$ km\,s$^{-1}$. The
uncertainty in $R\sin i$ is estimated from the uncertainty in $v\sin
i$, which is much greater than the uncertainty in $P$, giving a {\it
  fractional} uncertainty in $R\sin i$ of $S_{v\sin i}/v\sin i$.
   For targets where this fractional uncertainty is greater than
0.3, upper limits to $R\sin i$ are calculated from $P$ and
the upper limit to $v\sin i$. 
Figure 3 shows $R\sin i$ versus $M_K$ and predicted model radii $R_m$
from the BHAC15 and Dartmouth evolutionary codes (Dotter et al. 2008)
for 625\,Myr solar metallicity isochrones.  These models are
almost indistinguishable at this age and in this luminosity range.
Figure 4a shows the same data normalised to the BHAC15 model radii as a
function of luminosity.  The ratio of projected radius to model radius
at the target luminosity, $r\sin i = R\sin i/R_m$, is referred to
hereafter as the ``normalised radius''. There are 48 targets that have
a fractional uncertainty in $r \sin i$ of $\le 0.3$. A further 12
targets have $r\sin i$ upper limits and are treated as left-censored
data.


A maximum likelihood method was used to determine the average 
over-radius $\rho=R/R_m$, relative to the BHAC15 model radii
as a function of luminosity, that best matches the observed data in
Fig.~4a. The method is described fully in Paper I, but in brief consisted
of using Monte Carlo realisations for each target that assume a random
alignment of spin axes. The uncertainties estimated for each target were
used along with a
treatment of unresolved binarity (described in Appendix 1) and an
appropriate treatment of left-censored data, to produce a probability
distribution of $r \sin i$ for each target given its luminosity and
period and a value of $\rho$.  The best-fitting value of $\rho$ was
determined by maximising the log-likelihood function summed over all
targets ($\ln \mathscr{L}$). The uncertainty in $\rho$ is estimated
from the standard deviation of the likelihood function.

Figure 4b shows the number density of targets versus
$r\sin i$ compared with what would be expected from a similar set of
measurements if the stars have random spin-axis alignment and radii as
predicted by the BHAC15 isochrone ($\rho=1$, shown as a dotted line). The solid
line shows the best-fitting model with $\rho = 1.07 \pm 0.03$ (see section~4).

\section{Results}
\subsection{Over-radius with no correction for binarity or metallicity}
Figure 5 shows the measured period versus luminosity for targets with
$\log L/L_{\odot}<-0.2$, indicating
those stars for which we obtained spectroscopy and those stars
which were included as part of the $r \sin i$ analysis (those with $(v \sin
i)_p > 8$ km\,s$^{-1}$). The data were split
into two luminosity bins for analysis, spanning the approximate mass range
$0.15< M/M_{\odot}<0.6$, with roughly equal numbers of targets
per bin. The parent sample of the lower luminosity bin contains
almost exclusively faster rotating
stars with $P<3$ days. The upper bin includes both stars in the fast
rotating "C sequence" and stars in transition between the faster "C
sequence" and slower "I sequence" defined for F-K stars by Barnes
(2007). Note however, that even the slowest rotating stars that contribute to
the $r \sin i$ analysis in both bins have $P< 2$ days. There are too few stars with $(v \sin i)_p
>8$\,km\,s$^{-1}$ to make a useful estimate of over-radius at higher masses/luminosities.

\begin{figure}
	\centering
	\includegraphics[width = 77mm]{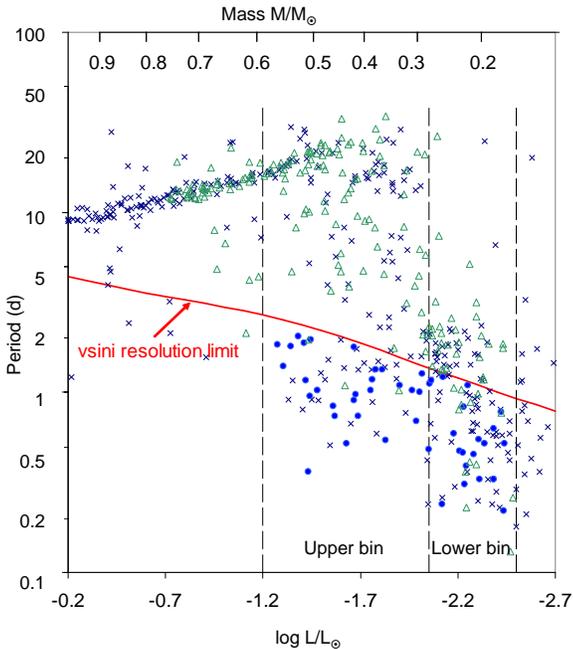}
	\caption{The rotation periods of low mass stars in
          Praesepe.  Triangles mark
          stars with measured spectra; Filled circles are the subset of
          stars with measured $r\sin i$ values with uncertainty $\leq
          30$ per cent. Crosses show other stars with measured periods
          reported by Douglas et al. (2017a) but {\it not} included in
          our observations. The red line marks the locus of stars with
          $(v\sin i)_p = 8$\,km\,s$^{-1}$; stars above this line are
          excluded from the $r \sin i$ analysis. Dashed vertical lines
          are boundaries that define the upper and lower luminosity
          bins.}
\label{fig5}
\end{figure}

The results of the maximum likelihood analysis are shown in Table~5. The
estimated over-radius relative to the {\it solar metallicity}
BHAC15 evolutionary model {\it with no correction made for binarity} is
$\rho = 1.07\pm 0.03$ (or an over-radius of $7 \pm 3$ per cent). The best fitting model
$r \sin i$ distribution is shown in Fig.~4. A model with a free value
of $\rho$ is preferred to one with $\rho$ =1. The difference in log
likelihood is 2.1 but with an additional free parameter; assuming
Wilks' theorem, a likelihood ratio test suggests that the null
hypothesis of no radius inflation can only be rejected with a p-value of
0.04 (i.e roughly 2-sigma). Results are also shown when the targets are
split into the two luminosity bins. A small radius inflation is
favoured over a model with no inflation for each bin, but only at a marginal
significance level.

\begin{table*}
\caption{The maximum likelihood value of over-radius $\rho$, for
  faster rotating low mass stars (with $(v \sin i)_p > 8$\, km\,s$^{-1}$)
   in Praesepe, relative to radii predicted for a 625\,Myr BHAC15
  model. $N_{\rm targ}$ is the number of targets included 
and $N_{\rm rsin i}$ is the number of those targets with measured
values of $r\sin i$. The column labeled $\Delta \ln \mathscr{L}_{\rm
  max}$ gives the increase in log-likelihood with respect to a model
(with no free parameters) that assumes $\rho$=1 and random spin-axis orientation. 
In the lower portion  results are
shown for a model with super solar metallicity and including a
treatment of unresolved binary stars (see Appendix~1). The
errors in $\rho$ are solely the statistical uncertainties. The final
row gives the results assuming the non-random spin-axis distribution
found by Kovacs (2018); in this case the change in log-likelihood is
with respect to our best fitting model with a random axis orientation,
that has $\rho=1.06 \pm 0.04$.} 
\begin{tabular}{lllllll} \hline
&	$N_{\rm targ}$	&
$N_{\rm rsini}$	&	$\log L/L_{\odot}$	&	$\overline{r\sin i}$
  &$\Delta \ln \mathscr{L}_{\rm max}$	&	$\rho$		\\\hline
\multicolumn{7}{l}{\underline{Fixed binary fraction = 0,  model [FEH]=0} }\\	
All targets 	&	60	&	48	&	-1.953	&	0.86	&	2.1	&	1.07$\pm$0.03\\
Lower luminosity bin 1	&	30	&	22	&	-2.268	&	0.916	&	1.2	&	1.09$\pm$0.05	\\
Upper luminosity bin 2	&	30	&	26	&	-1.639	&	0.813	&	1.0	&	1.06$\pm$0.04	\\
\multicolumn{7}{l}{\underline{Variable binary fraction,  model [Fe/H]=0.156}}\\														
All targets	                                         &	60	&	48	&	-1.953	&	0.860	&	1.1	&	1.06$\pm$0.04	\\
Lower luminosity bin 1	                             &	30	&	22	&	-2.268	&	0.916	&	1.2	&	1.09$\pm$0.05	\\
Upper luminosity bin 2																&	30	&	26	&	-1.639	&	0.813	&	0.3 &	1.03$\pm$0.05	\\
Slower rotators (bin 1 P$>$0.5\,d, bin 2 P$>$1.1\,d)	&	30	&	22	&	-1.948	&	1.007	&	3.8	&	1.14$\pm$0.04		\\
Faster rotators (bin 1 P$<$0.5\,d, bin 2 P$<$1.1\,d)	&	30	&	26	&	-1.959	&	0.737	&	0.3	&	1.02$\pm$0.04		\\
Higher amplitude light curves(amp $>$ 0.038 mag)	       &	31	&	26	&	-2.006	&	0.948	&	3.8	&	1.12$\pm$0.04	\\
Lower amplitude light curves (amp $<$ 0.038 mag)	        &	29	&	22	&	-1.897	&	0.757	&	2.2	&	0.96$\pm$0.03	\\
All targets - partial alignment of stellar axes (see Section 5.1)	&	60	&	48	&	-1.953	&	0.86	&	-21.8	&	0.97$\pm$0.03 \\\hline
\end{tabular}
\label{over-radii}
\end{table*}

\subsection{The binary fraction of targets with measured $r\sin i$}

The presence of unresolved binaries in the sample with
measured $r\sin i$ affects the inferred $\rho$ in two
ways (see Paper I).
First, the CCF may be broadened by any unresolved velocity
difference between the two components that contribute to the spectrum, 
leading to an over-estimate of $r\sin i$. Second, because
the system luminosity is larger than that of the primary
alone, the model radius is systematically
over-estimated in binary systems and hence $r \sin i$ is under-estimated.
Appendix 1 describes a detailed simulation that is used to provide a correction
for these effects, but this requires an estimate of the fraction of
unresolved binaries in the population.

Figure 6a shows luminosity versus $(V-K)_0$ for the
Praesepe targets. The faster rotating stars,
those with measured $r\sin i$ values, appear to lie
redward of other Praesepe targets on average, suggesting
that the sample used to evaluate $\rho$ may contain a larger proportion
of unresolved binary systems than the cluster as a whole. This is not the only possible explanation
of a bias in $V-K$, for example in section 5.2 the possibility that
large proportions of the stellar surfaces are covered by cool
starspots is discussed, but it would be consistent with the
conclusion of Douglas et al. (2017) that a high proportion of faster
rotating Praesepe stars with $M>0.3 M_{\odot}$ are probable binaries
and also has the largest effect on our inference of the over-radius.

Figure 6b shows the offset in $(V-K)_0$ as a function of luminosity, 
relative to an average value defined by cubic polynomial fits to the
data in each luminosity bin. Also shown in Fig.~6b is the distribution of
these offsets for a simulated population with a binary fraction,
$B_f=0.30$ (Duch{\^e}ne \& Kraus 2013), and a Gaussian uncertainty
of 0.2 mag in $(V-K)_0$. The simulation draws from a uniform distribution of
mass ratios between 0.1 and 1 for the binary stars (Raghavan et
al. 2010) and uses the BHAC15 model to estimate colour at a given mass.

\begin{figure*}
	\begin{minipage}[t]{0.98\textwidth}
	\includegraphics[width = 180mm]{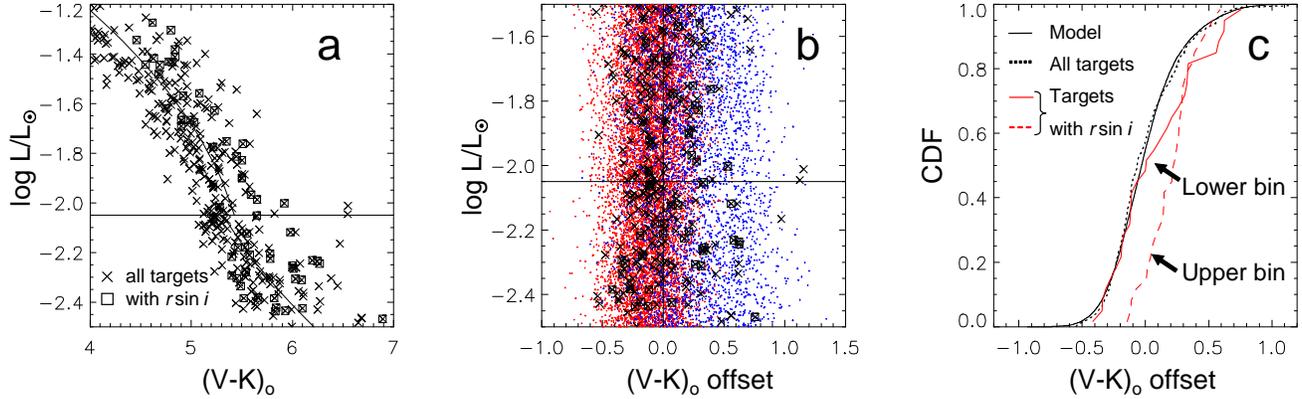}
	\end{minipage}
	\caption{Plot (a) shows $(V-K)_0$ for the Praesepe stars with measured
          periods in the upper and lower luminosity bins (delineated by
          the horizontal line). Squares are
          stars with a measured $r\sin i$, crosses are other targets, the
          trend lines are cubic polynomial fits to the data in each
          luminosity bin. Plot (b) shows the offset in $(V-K)_0$ relative to the
          trend lines in plot (a). Dots show a Monte-Carlo model distribution of
          single stars (red) and binary stars (blue) for a model with
          a binary fraction of 0.3 and a measurement uncertainty of 0.2
          in $(V-K)_0$ (see Section 4.2). Plot (c) compares the
          cumulative distribution functions (CDFs) of the
          $(V-K)_0$ offset shown in plot (b) for all data (black dotted line) and 
          and for just the data with measured $r\sin i$ 
          (red solid and dashed lines for the lower and
          upper luminosity bin respectively) to the CDF of the
          Monte-Carlo model data shown as a back solid line.}
	\label{fig6}
\end{figure*}

Figure 6c compares the cumulative distribution functions (CDFs) of the
offsets in Fig.~6b.  The CDF for all
targets with measured periods is well-described by the Monte-Carlo
simulation. The CDFs for the subsets of stars with measured values of
$r\sin i$ are poorly described by the simulation with
$B_f=0.30$. A better estimate of $B_f$ can be obtained by using Monte
Carlo simulations to determine the probability of a target being a
binary as a function of $(V-K)_0$. This yields an average $B_f= 0.47$
for the targets with $r\sin i$ in the upper luminosity bin, and
$B_f=0.38$ for those in the lower luminosity bin. Whilst this is only
an approximate method for estimating the binary fraction for the subset
of data with measured $r\sin i$, it is acceptable when used to assess the
effects of binarity on $\rho$ which is itself a relatively small correction.

\subsection{The effects of binarity and metallicity on over-radius}

Appendix 1 describes the Monte-Carlo model used to estimate the effects
of unresolved binarity on the inferred $\rho$. Including the effects of
binarity, assuming the binary frequencies estimated in the previous
subsection, a period distribution appropriate for field stars and a
flat mass-ratio distribution {\it increases} the estimated $\rho$ in
the Praesepe sample by just 1 per cent. If some of the redward
displacements seen in Fig.~6 are in fact caused by spottedness rather
than binarity, then the binary fraction will be lower and the influence
on $\rho$ will be smaller.
    
Praesepe has a super-solar metallicity; [Fe/H]$=+0.156\pm 0.004$
(Cummings et al. 2017). In general, evolutionary models (Baraffe et
al. 1998, Dotter et al. 2008) show an increase in radius at fixed
luminosity with increasing metallicity. Since BHAC15 models at
alternative metallicities are not available, the effect of metallicity
on the model radii is estimated from the Dartmouth model radii for
[Fe/H]=0 and [Fe/H]=0.16 (Dotter et al. 2008). These models showed an average
increase in radius of 2.0 per cent at the higher metallicity (at an age
of 600\,Myr, over the range $-2.5< \log L/L_{\odot} < -0.6$). At the
measured Praesepe metallicity this
corresponds to a 2 per cent increase in the model radius over the
solar metallicity values and
hence a uniform  {\it decrease} in $\rho$ of $-0.02$.

Table 5 shows the estimated value of $\rho$ for fast rotating stars in
Praesepe after accounting for the effects of both binarity and super-solar
metallicity, assuming a binary fraction of 38 per cent for targets in
the lower luminosity bin and 47 per cent for targets in the upper
bin. The net effect is to decrease the over-radius by just $\sim 1$ per cent
compared with the value inferred assuming a solar metallicity and no
unresolved binaries. The effects on the lower and upper luminosity bins
are slightly different. At lower luminosities $\rho$ is
almost unchanged, whilst for the upper luminosity bin
$\rho$ is decreased by $-0.03$.

\subsection{Effects of light curve amplitude and rotation rate.}
Table~5 also compares $\rho$ determined for subsets of the fastest and more
slowly rotating stars and for targets showing higher and lower
levels of light curve modulation (reported in Douglas et al. 2017). In
each case, the samples were chosen by splitting each of the two
luminosity bins at their median rotation period and light curve
amplitude respectively. The
slower rotating stars have $\rho = 1.14 \pm 0.04$, a much more
significant result, with a p-value of 0.002 with respect to a model
with no inflation. On the contrary, the fastest rotating stars show
little evidence for inflation and an over-radius that is significantly lower.
Targets that exhibited a higher level of light curve
modulation (above the median value of 0.038 mag) also display evidence for
significant inflation (a p-value of 0.001) and have a $\rho$ that is
is higher than targets with lower levels of light curve
modulation, which do not evidence significant radius inflation.

The difference in $\rho$ between the low- and high-amplitude rotational
modulation subsets is expected, since there should be a strong 
positive correlation between
modulation amplitude and $\sin i$  --
analysing these subsets with a random $\sin i$ should lead to an
underestimated or overestimated $\rho$ respectively.  However, there is no
similar bias that can explain why the inferred $\rho$ of the faster and
slower rotating subsets should be different. Whilst the uncertainties in the
estimated value of $\rho$ for individual bins are large due to the small
sample sizes, the trend of higher $\rho$ for slower rotators and for
stars exhibiting higher levels of light curve modulation are the same
as those seen in the larger sample of Pleiades targets in Paper I.

\subsection{Systematic uncertainties in the over-radius}
Table 5 gives {\it statistical} uncertainties in $\rho$ 
based on the standard deviation of the likelihood function, which in
turn are chiefly dependent on the number of targets in the analysed
sample. Other systematic uncertainties were discussed in Paper I
including uncertainty in cluster properties, the effect of surface
differential rotation (SDR) and bias due to selection effects in the
period data.
\begin{figure*}
	\centering
	\includegraphics[width = 165mm]{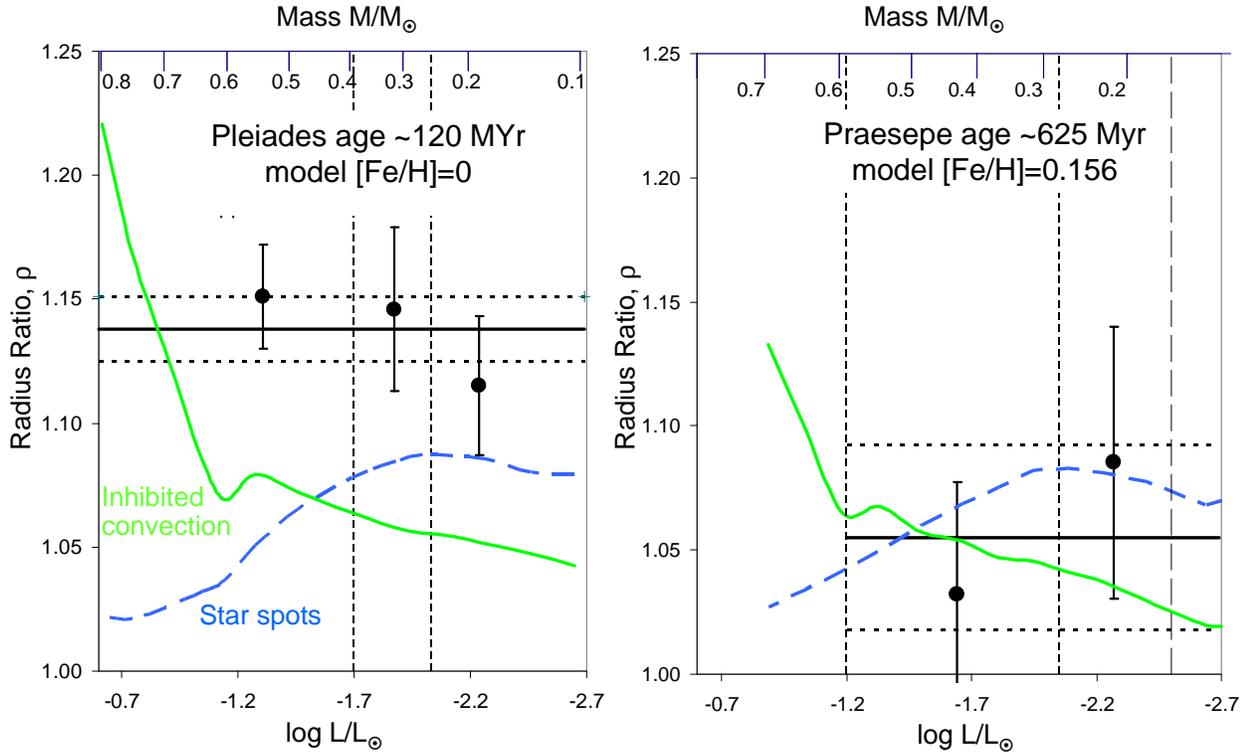}
	\caption{The estimated over-radius of fast rotating low mass
          stars in the Pleiades and Praesepe relative to the 
          model predictions of
          BHAC15 at the cluster age. Results in Plot (a) are reproduced
          from Paper I; results in Plot (b) are from results reported
          in this paper. Horizontal lines show the mean over-radius,
          for all the data, with dashed lines indicating the
          1-sigma confidence interval. The individual points with error
          bars show the mean over-radius and 
          uncertainties for stars in luminosity/mass bins. The mass
          scale at the top of the plot is based on the same BHAC15
          model. Green solid lines shows the predicted effect of radius
          inflation due to magnetic inhibition of convection (Feiden et
          al. 2015); the blue dashed lines show the predicted
          effect of starpots with an effective dark spot coverage of
          $\beta =0.16$ (see section 5.2).}
\label{fig7}
\end{figure*}

For Praesepe the uncertainties in $\rho$ due to uncertainties in age
and reddening are negligible. The uncertainty in distance modulus
$(M-m)_0$ is $\pm 0.04$\,mag, 
assuming a possible 0.1 mas remaining 
systematic uncertainty in the mean parallax (Gaia collaboration 2018). This yields an
additional uncertainty of $\mp 0.014$ in $\rho$. Whilst the
estimated metallicity of the cluster is quite consistent between
studies (Cummings et al. 2017, Yang et al. 2015, Boesgaard et al. 2013)
it is less certain how super-solar metallicity affects stellar radius
at a fixed luminosity. Similarly the correction made for binarity in
Appendix 1 is approximate. Making a conservative assumption that the
corrections made for metallicity and binarity are each accurate to
$\pm$50 per cent gives uncertainties in $\rho$ of only $\pm 0.01$ and
$\pm 0.008$ respectively. It was shown in Paper I that SDR had little effect
($<1$ per cent) on the measured periods of fast-rotating low mass stars
and it is neglected here.

Any bias due to selection effects in period measurement depends on the
completeness of the period data; i.e. whether the sample with
measured periods are 
representative or preferentially excludes targets with low
inclinations.  Selecting only the 50 per cent of Praesepe targets with
the largest light curve amplitude would increase the estimated value of
$\rho$ by $+0.06$ compared to an analysis of the entire
sample (see Table 5), presumably because this sample contains stars
with that are heavily biased towards higher $\sin i$. 
This is an extreme case; Douglas et al. (2017) reported 
periods for 86 per cent of the Praesepe targets in the luminosity range
considered here. Even if
the missing 14 per cent of targets without periods were exclusively at the low end of the
$\sin i$ distribution, the inferred $\rho$
would be over-estimated by just $+0.01$.

  A final source of systematic error would be if the rapid rotation
  of the stars caused them to be sufficiently oblate to
  compromise either the calculated model radii or the $v \sin i$
  measurements. An assessment of this effect can be made using
  equation~46 and the coefficients in table~7  provided by
  Chandrasekhar (1933) for the equilibrium configuration of rotating
  polytropes. Taking an extreme example of a
  fully convective
  $0.3M_{\odot}$ star with a  rotation period of 0.2\,d (see Fig.~5)
  and a central density of $\sim 100$ g\,cm$^{-3}$, the predicted
  increase in equatorial radius is just 1.1 per cent. This will be an
  upper limit to the consequent increase in measured $v \sin i$, since
  the radius at higher latitudes is smaller, so we
  conclude that this effect can be neglected.

Thus overall, any additional systematic uncertainty is likely to be
$\sim 2$ per cent and smaller than the existing statistical
uncertainties listed in Table~5.

\section{Discussion}

\subsection{The dependence of radius inflation on age, mass and rotation}

The main aim of this investigation was to measure the radii of low-mass
stars in Praesepe and to compare them with the predictions of
evolutionary models and also to compare any over-radius with that
estimated for stars of similar mass in the Pleiades using the same
analysis methods and very similar data.

The inferred over-radius for this sample of relatively fast-rotating
stars in Praesepe is $6 \pm 4$ percent (after correction for
metallicity and binarity, see Table~5) with an additional $\sim 2$
percent of systematic uncertainty. The difference in log-likelihood
between a model with no over-radius and the best-fitting model is
however not very large; although our measurement is consistent with a
small inflation, we cannot be certain at the $>2$-sigma level that
these stars are bigger (at a given luminosity) than predicted by the
most commonly used low-mass evolutionary models that do not account for magnetic
effects (Dotter et al. 2008; BHAC15).

Figure~7 makes a comparison of the over-radii measured for the Pleiades
(from table~5 in Paper I) and Praesepe. The predicted radii
of stars in the Pleiades at 120\,Myr are slightly larger than those of similar
luminosity in Praesepe at 625\,Myr due to the completion of PMS
contraction in that interval. Figure~7 has already taken this into account -- the
over-radii are calculated with respect to models at the appropriate age.
In Fig.~7 the mean over-radius of $\rho = 1.14 \pm 0.02$ in the Pleiades is greater than
Praesepe at the 2-sigma level; the inclusion or not of the increased metallicity of Praesepe
and the effects of unresolved binarity (see Table 5) changes this difference by
only $\sim 1$ percent.
The stars in Praesepe are rotating somewhat slower on average than the
Pleiades targets from Paper I, however the more slowly rotating half of
the Pleiades targets, which are much more comparable to the present Praesepe
sample (see table~5 in Paper I), have an over-radius of $\rho = 1.16
\pm 0.03$ and so this does not change any of the discussion above.

\begin{figure}
	\centering
	\includegraphics[width = 85mm]{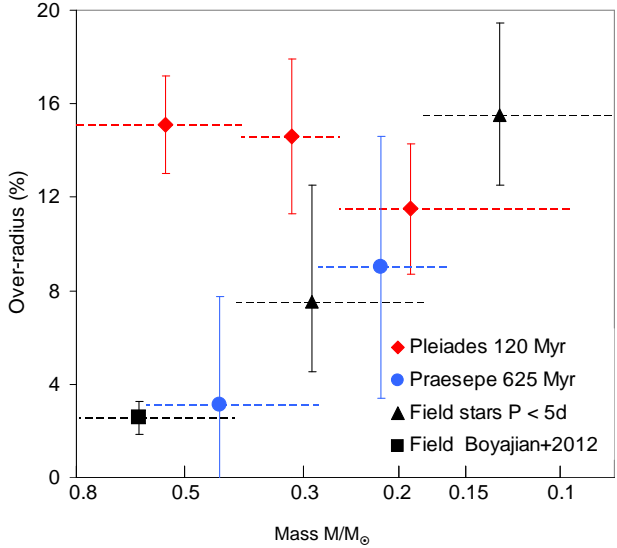}
	\caption{Composite plot showing the mean over-radius measured on samples of low mass stars from four studies,	(a) over-radii of stars in the Pleiades (paper I), (b) over-radii in the Praesepe (this paper), (c) over-radii of faster rotating field stars (Kesseli et al. 2018) and (d) over-radii of slowly rotating field stars (from paper I based on radii reported by Boyajian et al. 2012) as a function of mass determined from stellar luminosities. Vertical error bars show 1$\sigma$ uncertainties in over-radius. Horizontal dashed lines show the mass range of binned data contributing to each point.} 
	\label{fig8}
\end{figure}

The final over-radius estimates of $14 \pm 2$ per cent for the Pleiades
and $6 \pm 4$ per cent in Praesepe can also be compared with
the over-radius of $13^{+3}_{-2.5}$ per cent with respect to the 
BHAC15 models reported by Kesseli et al. (2018)
for a sample of field M-dwarfs with rotation periods less than 5
days. Kesseli et al. (2018) used a very similar analysis technique, but
our focus on samples of stars in clusters removes uncertainties associated with
the age and metallicity of a field star sample.

Figure 8 collates these over-radius measurements into one plot
and adds higher mass stars with interferometric radii from Boyajian et
al. (2012). The masses here are estimated from the 
BHAC15 models, assuming an age of 5\,Gyr for the field stars,
and the over-radii are with respect to BHAC15 predictions for the
Pleiades, Praesepe and interferometric samples, and with respect to the
Dartmouth models for the sample from from Kesseli et al. (2018).
Kesseli et al. (2018) reported that the lowest mass stars in their sample
(estimated to be $\sim 0.1 M_{\odot}$ using non-magnetic evolutionary
models) are marginally more inflated ($15.5^{+4}_{-3}$ per cent) with respect to the model
predictions than stars of higher mass ($7.5^{+5}_{-3}$ per cent inflation for stars
with $0.18 < M/M_{\odot}<0.4$). Our lowest mass targets in the Pleiades
and Praesepe are intermediate to these samples in both mass and
estimated over-radius. Figure~8 suggests, albeit strongly based on the Pleiades
data, that the decrease in $\rho$ between the age of the Pleiades and the
Praesepe/field-star samples is driven by changes for stars with $M > 0.25 M_{\odot}$, but
that there may be no decrease for lower mass stars. 

It should be noted that the masses quoted here are derived from
  standard model-dependent mass-luminosity relationships. In section~5.2 we
  consider whether models featuring magnetic inhibition of convection
  or starspots may provide better descriptions of the data and this may
  change the mapping of luminosity to mass for different groups of
  stars. However, {\it all} of the stars in the samples shown in Fig.~8
  are strongly magnetically active apart from the sample of nearby
  field stars with interferometric radii. The increase in inferred
  mass, at a given luminosity and at ages $\geq 100$\,Myr, compared with
  non-magnetic models 
  is at most $+0.05 M_{\odot}$ for any of the
  magnetic models considered in section~5.2 and so should not greatly compromise
  any comparisons between samples divided on
  estimated mass bins that are much
  broader than this.

Kesseli et al. (2018) also reported that they find no significant
difference in inflation for the fast versus slow rotators, although the
comparison is between inflation determined from their
fast-rotating sample versus the radii for slower rotating stars
determined by other means -- either interferometrically (for which there are only a couple
of examples at a comparable mass) and for the components of one slowly
rotating eclipsing binary system. On the contrary we find, by dividing
our sample of stars at their median period,
that the fastest rotating stars in our sample, are {\it
  less} inflated than those with slower rotation by
12 $\pm$6 per cent (see Table 5), although all of these stars are
easily rapidly rotating enough to have saturated levels of magnetic
activity. For example, the stars in our Praesepe sample have $(V-K)_0>4$ (see Fig.~6),
and corresponding convective turnover times of $>50$ days (e.g. see
Wright et al. 2018). Magnetic saturation as judged by coronal X-ray and
chromospheric emission appears to set in below Rossby numbers (rotation
period/convective turnover time) of 0.1 at
all the spectral types considered here (e.g. Jeffries et al. 2011;
Wright et al. 2011, 2018; Newton et al. 2017) and thus all stars with
$P<5$ days are in the saturated regime, which includes all targets used
for the $r\sin i$ determinations (see Fig.~5).

In Paper I we suggested that it is the increased radii of these stars
that could be {\it responsible} for their slower rotation rates, rather
than the other way around. Angular momentum losses via a magnetically
coupled wind may be strongly radius-dependent (Reiners \& Mohanty
2012; Matt et al. 2015), or may increase for some other reason in the more inflated stars
-- perhaps a change in the magnetic field configuration (e.g. Garraffo,
Drake \& Cohen 2016). If the
timescale for significant angular momentum loss becomes shorter than
the timescale on which the stellar radius can change, then any decline
of observed over-radius with age could be explained in terms of a
spread in radii. The larger stars would spin down more rapidly, and
since the sample of stars for which we can measure $r \sin i$ is
strongly biased towards the most rapid rotators, then the more inflated
stars would preferentially be excluded in older samples.  The
mass-dependence of the over-radii seen in Fig.~8 could then merely be a
consequence of the much longer spindown timescales in lower mass stars
-- the majority of stars in the lowest luminosity bin of Praesepe are
still fast rotators for which a $r \sin i$ measurement is possible,
whereas in the higher luminosity bin, more than half have spun down to
rotation levels where $r \sin i$ could not be measured in our
spectroscopy (see Fig.~5).

\subsection{Comparison with magnetic models}

A significant over-radius with respect to ``standard'' evolutionary
models has been identified in fast-rotating M-dwarfs in the Pleiades
(Paper I) and now (marginally) for the lowest mass stars in Praesepe
(this paper) and in a field star sample of very low-mass stars (Kesseli
et al. 2018). However, where the data exists (primarily at the upper
end of our considered mass range) there is a much better agreement
between the same evolutionary models and interferometric radii for
field stars of similar mass but lower levels of magnetic activity
(e.g. Boyajian et al. 2012), suggesting that magnetic activity may be
responsible for the discrepancies. Two flavours of models incorporating
the effects of magnetic fields are provided by: (i) magnetic inhibition
of convective flux (e.g. Feiden \& Chaboyer 2012, 2014) or the blocking
of radiative flux from the stellar surface by cool, magnetic starspots
(Jackson \& Jeffries 2014a; Somers \& Pinsonneault 2015a,b).

Magnetic inhibition of convection should become less effective at lower
masses as the stars become fully convective, and the effect is also
predicted to weaken for older stars that have become established on the
main sequence compared with PMS stars (see for example Feiden \&
Chaboyer 2014; Feiden et al. 2015). Both of these effects are apparent
in Fig.~7, where we plot the over-radius (compared to standard models)
of a 120 Myr and 625\,Myr isochrone from magnetic models that implement
magnetic inhibition of convection via a rotational dynamo, with a fixed surface field of 2.5\,kG
(Feiden et al. 2015).
The data from Praesepe alone are possibly
consistent with this model.  However, the model is inconsistent with
the additional data provided by the Pleiades and the older field star
sample of Kesseli et al. (2018), both of which show significantly
larger over-radii than predicted by the magnetic models, and in the
case of the field stars (and perhaps Praesepe too), an over-radius that
increases with decreasing mass (see Fig.~8). 

A field of 2.5\,kG
  represents an approximate equipartition value for these low mass
  stars, where the thermal and magnetic pressures in the photosphere
  are roughly equal. Some recent measurements using the Zeeman effect
  suggest that average surface magnetic fields may reach strengths of
  7\,kG,
  but only in low-mass, fast-rotating, fully-convective stars (Shulyak et
  al. 2017). The models of Feiden et al. (2015) suggest (see their fig.~2)
  that the inflationary effect of magnetic fields scales more
  steeply than linearly and so the mass- and age-dependence of radius
  inflation might be explained if the lowest mass stars in our Pleiades
  and Praesepe samples, and the most active fully-convective field stars, 
  host super-equipartition fields that persist for billons of years.

A further alternative could be provided by dark starspots. A large starspot
coverage reduces the radiative flux from the surface and results in an
over-radius with respect to an immaculate star of the same mass and age.
The level of inflation depends on the fraction $\beta$ of the flux
blocked by spots. A second order effect is that the colours and
bolometric correction of the star will be changed depending on both the
spot coverage and the ratio of spotted to unspotted surface temperature
(e.g. Jackson \& Jeffries 2014b). In Paper I we were able to take
advantage of spectroscopic observations of the Pleiades low-mass stars,
which were modelled with two temperature components by Fang et
al. (2016). This revealed a range of $\beta$ values with a mean of 0.16
for the sample of fast-rotating stars considered there. Figure~7 shows
the predicted over-radius due to spots with $\beta=0.16$, interpolated
from the models of Somers \& Pinsonneault (2015a). The spot model with
this value of $\beta$ provides a good match to the Praesepe data but is
incapable of producing the larger over-radii in the Pleiades,
especially at the higher masses/luminosities, without increasing the
adopted value of $\beta$ (or including some additional element of inflation due to
suppression of convection).

Unfortunately, there are no spectroscopically determined $\beta$ values
for our Praesepe targets, so it is unknown whether the level of spot
coverage changes between the ages of Praesepe and the Pleiades or
for older field stars. The median rotation period of the stars
certainly does increase, but all stars contributing to the estimate of
$\rho$ rotate fast enough to be at saturated magnetic activity levels.
As a proxy, we can look at the distribution of light curve amplitudes. 
For spot coverages less than 50
per cent, it is likely that spot coverage and light curve amplitude
will be positively correlated, with amplitude roughly proportional to
$\beta^{1/2}$ for spots of a given size and temperature (Jackson \&
Jeffries 2013), although this relationship would be complicated by any changes in
the latitudinal distribution of spots or changes in the mean spot
size. 

Figure~9 shows a comparison of the cumulative
distribution of spot amplitude (defined in the same way) for Praesepe
versus the Pleiades, which considers {\it only} the stars used in the
respective $r \sin i$ analyses, split at their median values 
into faster and slower rotating
subsets. The slower rotating half of the Praesepe sample have a larger mean
light curve amplitude than the faster-rotating half. A
Kolmogorov-Smirnov test suggests a 99 per cent significant difference
between the cumulative distributions. This might fit in with an
explanation involving radius inflation due to starspots but there is no
straightforward way to translate this into a difference in $\beta$. The
difference between the faster and slower rotating samples in the
Pleiades is in the same direction, but is much less significant. There
is also no significant decline in light curve amplitudes between
the faster rotating subsets in the Pleiades and Praesepe to accompany
the observed decrease in $\rho$.

In conclusion, the results are inconclusive. The mass-dependence that
we identify seems inconsistent with the magnetically inhibited
convection models with equipartition magnetic fields at the
  surface. Some fast-rotating, low-mass, fully-convective field stars have
  been found to host stronger magnetic fields and such fields might be
  capable of providing the observed mass-dependence. A mass-dependent
  starspot coverage could also provide this mass-dependence, but 
there is no strong evidence for a
decrease in spot coverage with age that might explain the reduced
over-radius among the higher mass stars in Fig.~8. These considerations
are complicated by the apparent rotation-dependence of the over-radius,
which could be a selection effect cause by the progressive spindown of
larger stars and their exclusion from our samples.

\begin{figure}
	\centering
	\includegraphics[width = 75mm]{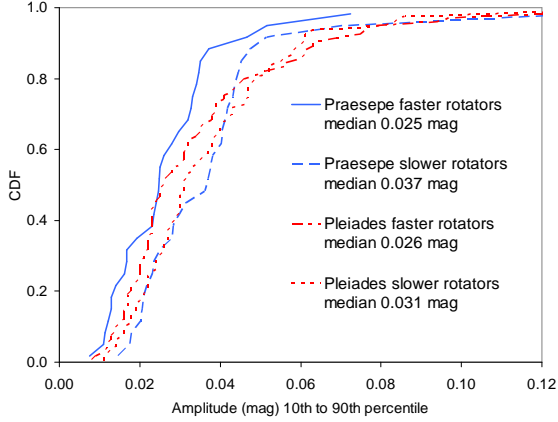}
	\caption{Cumulative distributions of light curve amplitude for stars with measure values of r$\sin i$ in Pleiades (Paper I) and Praesepe (this paper) split into fast and slower rotating subsets. } 
	\label{fig9}
\end{figure}
\subsection{Non-random axis orientation}

The analysis so far has made the assumption that there is no
preferred orientation of the spin-axes. Recent work by Kovacs (2018)
has questioned this assumption, specifically with regard to the
Praesepe cluster. Using quite similar techniques to those discussed
here -- rotation periods from Kepler K2 data and literature $v \sin i$
values for 120 F-K dwarfs from Mermilliod, Mayor \& Udry (2009) --
Kovacs found that for this higher mass stellar sample, the $r \sin i$
distribution was better fitted with a model that had no over-radius but
a narrower, aligned $\sin i$ distribution that was characterised with
spin-axes at an average inclination of $76^{\circ} \pm 14^{\circ}$ and
distributed over a cone with a half opening angle of $47^{\circ} \pm
24^{\circ}$.

\begin{figure}
	\centering
	\includegraphics[width = 80mm]{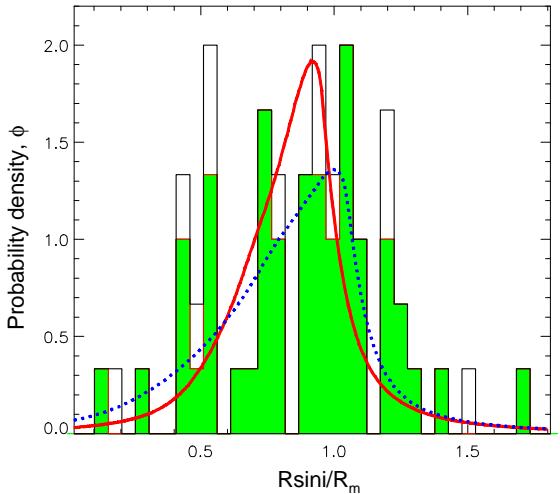}
	\caption{Comparison of the the measured distribution of $r\sin i$ as a function of luminosity with the best 
	fit model distributions found using the maximum likelihood model. The dotted line shows results for a random distribution of stellar spin-axis orientation and $\rho = 1.06 \pm 0.04$. The solid line shows results for a partially aligned distribution with an average inclination of $76^{\circ}$ distributed over a cone with a half opening angle of $\lambda$=$47^{\circ}$ and $\rho$=0.97$\pm$0.03.}
	\label{fig10}
\end{figure}

We have investigated whether our data are consistent with this
model. The results are shown in Table~5 and Fig.~10. The effect of
restricting the spin-axes to a cone with this relatively high inclination
angle, increases the mean value of $\sin i$ to 0.90. As a result, a
much lower value of $\rho$ is required to fit the data and we find
$\rho = 0.97 \pm 0.03$, consistent with no over-radius at all.
However, the fit to the data is much worse; the narrower range of
predicted $r \sin i$ produced by this model is a poor match to the observed
distribution. The difference in log
likelihood of $-21.8$ (see Table~5, with the same number of
data points and free parameters) decisively favours the model with a 
random axis orientation. Indeed the aligned model is disfavoured even
versus a model with $\rho=1$ and a random axis orientation ($\Delta \ln
\mathscr{L}$=-19.7).

In Paper I we conducted tests with a range of mean axis inclinations
and cone opening angles for the Pleiades sample, concluding that while
alignments were possible if the mean inclination angle was
$<45^{\circ}$, that these would require much higher values of $\rho$ to
compensate. Non-random axis orientations with higher mean inclinations were poor
fits to the data and were ruled out. We see the same effect in Praesepe for the particular 
case of an average inclination of $76^{\circ}$ (and $\lambda=47^{\circ}$).
In conclusion we can rule out the non-random
axis distribution proposed by Kovacs (2018) or any strong alignment
with a large mean axis inclination in Praesepe, but weak alignments or strong
alignments with a low mean axis inclination (and much larger
over-radius) are still possible. It is worth noting however that
Corsaro et al. (2017) have suggested, on the basis of numerical
simulations, that any residual alignment resulting from the initial
angular momentum and cluster formation process, might only be apparent
in the higher mass stars of a cluster.

\section{Summary}

Published rotation periods from Kepler K2 and new
measurements of rotational broadening have been combined to estimate
projected radii for a set of stars in the Praesepe cluster.
Adopting a random distribution of spin-axis orientation and using a maximum
likelihood technique, the average radius of
these fast-rotating ($P \sim$ 2 days or less), low-mass
($0.15<M/M_{\odot}<0.6$) stars is inferred to be 7$\pm$4 per cent
higher at a given luminosity than predicted by the
solar-metallicity isochrones of Baraffe et al. (2015) and Dotter et al. (2008).
Allowing for unresolved binarity
and the super-solar metallicity of Praesepe reduces the estimated
over-radius to $6\pm 4$ per cent and perhaps consistent with no
inflation at all.


The average over-radius in Praesepe is lower (at a 2-sigma 
significance level) than the $14 \pm 2$ per
cent measured for a larger sample of targets with similar mass in the younger
Pleiades using similar techniques (Jackson et al. 2018). 
Most of this evolution appears to occur at
the higher mass end of the samples, in agreement with the results found
for low-mass field stars using similar techniques (Kesseli et
al. 2018). The trend of more inflation at lower masses goes against the
predictions of models incorporating inhibition of convection by
dynamo-generated interior magnetic fields unless the fields are much
stronger than equipartition values in lower mass stars. An interpretation involving
extensive coverage by starspots may be more consistent with the mass
dependence, but there is no evidence that spot coverage declines
between the Pleiades and Praesepe in the studied samples. Although the Praesepe targets are slower rotating on average
than the previously studied Pleiades, a comparison with the more slowly
rotating half of the Pleiades sample does not alter the conclusions
above, and measurements for both clusters are based 
exclusively on stars that rotate fast enough to have saturated
levels of magnetic activity.

Another notable feature, that was also observed in the
Pleiades, is that the fastest rotating stars in our sample are the {\it
  least} inflated with respect to the models. An interpretation of this
could be that angular momentum loss is strongly radius dependent (as
suggested by Reiners \& Mohanty 2012 and Matt et al. 2015) and that larger stars are spun
down more rapidly and hence progressively disappear from the analysed samples because
they do not spin fast enough for their rotational broadening to be measured.

The main analysis assumes that the orientation of spin axes in the
parent sample is random. Recent work on higher mass stars in Praesepe by Kovacs
(2018) has suggested a systematic alignment of the spin axes. Such an
alignment is ruled out by the relatively broad distribution of
projected radii in our sample. Weaker alignments, or alignments
with a low mean inclination angle are still possible, but would mean
that the over-radius we have found is an under-estimate.

\bibliographystyle{mn2e} 
\bibliography{references}

\section*{Acknowledgments}

Data presented herein were obtained at the WIYN 3.5m Observatory from
telescope time allocated to NN-EXPLORE through (a) the scientific
partnership of the National Aeronautics and Space Administration, the
National Science Foundation, and the National Optical Astronomy
Observatory, and (b) Indiana University's share of time on the WIYN
3.5-m. This work was supported by a NASA WIYN PI Data Award,
administered by the NASA Exoplanet Science Institute, though JPL RSA \#
1560105.  RJJ and RDJ also wish to thank the UK Science and Technology
Facilities Council for financial support.

\appendix

\begin{figure*}
            \centering
            \begin{minipage}[t]{0.98\textwidth}
            \includegraphics[width = 180mm]{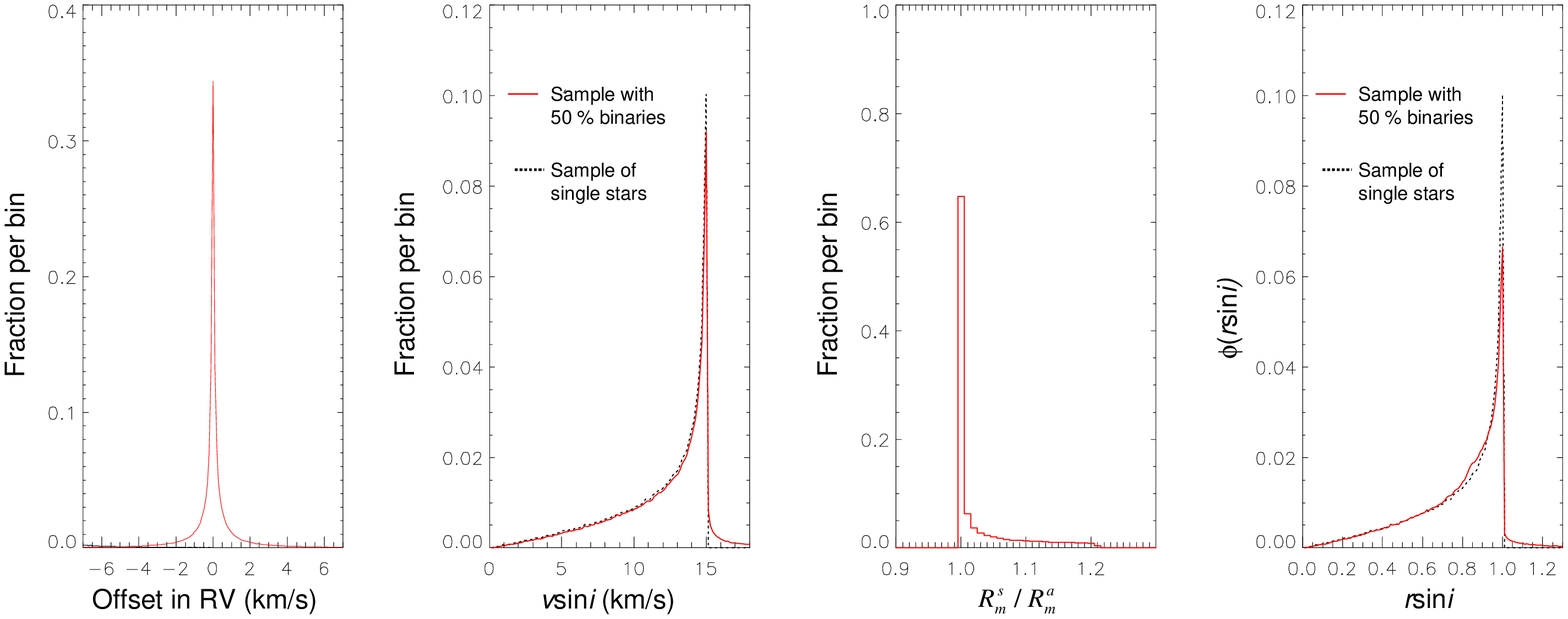}
            \end{minipage}
\caption{An example of the predicted effects of binarity on a population of stars
  with mass (of the primary star if binary) $0.5\,M_{\odot}$, an equatorial
  velocity of 15\,km\,s$^{-1}$ and a binary fraction 0.5 (see Section
  A1). (a) The distribution of primary star RV relative to
  the centre of mass.
  (b) The distribution of measured $v\sin i$ for the sample containing
  unresolved binaries and the equivalent distribution for single stars
  (dotted line).
  (c) The distribution of model radii $R_m^{s}$ determined
  from the system luminosity relative to the radius of single
  stars, $R_m^a$ with similar luminosity. (d) The combined effect of the changes in
  measured $v\sin i$ and predicted $R_m$ on the probability density function of $r\sin i$
  for a population containing unresolved binary systems compared with
  the equivalent distribution for a population of single stars (dotted line).}
            \label{fig11}
\end{figure*}

\section{Binary star simulations}

The normalised radius, $r\sin i$ is the ratio of the projected radius,
$R\sin i$ to the model radius $R_m$ at the measured luminosity, $L$.  
In the case of unresolved binary stars the measured
values of both $R\sin i$ and $R_m$ are increased above the true values
of the primary star (see Paper I). Depending on the binary frequency
and binary properties
this can either increase or decrease the estimated value of the over-radius
$\rho$.

\subsection{Modelling binarity}

Our method for modelling the effects of unresolved binarity on the CCF follows
that described in Jackson et al. (2016) and depends on the difference in
$RV$ between the primary and secondary and their relative
contribution to the observed spectrum. The resultant CCF is modelled as
the sum of two Gaussian profiles; the first represents the primary star with central velocity $RV+RV_a$ and
FWHM$_a$, where $RV$ is the systemic velocity and $RV_a$ is the
line-of-sight velocity of the primary relative to the centre of mass;
the second represents the secondary with velocity $RV-RV_a/q$ and
FWHM$_b$, where $q$ is the binary mass ratio. The resultant FWHM$_s$
of a single Gaussian fitted to this sum can be estimated as a function
of $RV_a$ and the ratio of flux contributions from the two stars (which depends on
$q$).

 
To a reasonable approximation the effect of rotational broadening is to
increase the width of the CCF as FWHM=FWHM$_0\sqrt{1+(v\sin i/C)^2}$
where FWHM$_0$ is the unbroadened width and $C$ is a constant that is 
proportional to the resolution of the spectrograph. Using this expression
the increase in the measured $(v \sin i)_s$ relative to the true $(v\sin
i)_a$ of the primary is, $C(\sqrt{{\rm FWHM_s}/{\rm FWHM}_0-1} -
\sqrt{{\rm FWHM}_a/{\rm FWHM}_0-1})$ and hence (from Eqn. 1) the
increase in inferred $R\sin i$, relative to that of the primary is
given by
\begin{equation}
\frac{(v\sin i)_s}{(v\sin i)_a} = \frac{\sqrt{{\rm FWHM_s}/{\rm FWHM}_0-1}}{\sqrt{{\rm FWHM}_a/{\rm FWHM}_0-1}}      
\end{equation}

A Monte Carlo model is used to calculate the $\it average$ effect of
binarity on $\rho$ as a function of target mass and $v\sin i$ assuming
a log normal distribution of binary periods (with $\overline{\log
  P}=5.03$ and $\sigma_{\log P}=2.28$) and a uniform distribution of
$q$ between 0.1 and 1 (Raghavan et al. 2010). Typical results are shown
in Fig.~A1 for a star of primary mass $M_a=0.5M_{\odot}$, a binary
fraction of 0.5 and a true equatorial velocity of $v=15$\,km\,s$^{-1}$.
Figure~A1a shows the distribution of $RV_a$ relative to the centre of mass
and Fig.~A1b shows the distribution of $(v\sin i)_s$ for a sample including
50 per cent unresolved binaries (those with $|RV_a - RV_b|
<10$\,km\,s$^{-1}$ for our spectrograph) compared to the distribution
of $v \sin i$ for single
stars. The net effect is to increase the average value of $v\sin i$ by
3.5 percent and produce a small tail of targets with a measured $v\sin i$ in
excess of the true equatorial velocity.

Figure~A1c shows the effect of binarity on $R_m$. For each binary in the
simulation we estimate the luminosity of the primary and the system
(from $M_a$ and $q$ using BHAC15 model) and use this to calculate
the increase in model radius, $R_m^{s}/R_m^a$. The distribution shows
a tail of stars with $R_m^{s}/R_m^a>1$, due to binaries with $q \sim
1$, which increases the average
value of $R_m$ by 2.5 per cent and hence decreases $r \sin i$. 
Figure~A1d shows the combined effects of
binarity on the base probability density function of $r\sin i$ used in
the maximum likelihood analysis. In the absence of binarity and
measurement uncertainty the base probability function follows the
dotted line (see Eqn. 5 for case of $\rho=1$). Binarity changes this
distribution producing both a tail stars with $\phi>1$ and a small
``bump'' at $r\sin i \sim 0.8$ due to binaries with $q \sim 1$.

\subsection{The effects of binarity on inferred over-radius}

The net effect of unresolved binarity on estimates of $\rho$ is not
straightforward. The two factors discussed above act in opposite directions and scale
differently with $v\sin i$. The increase in $\rho$ due to unresolved
binarity increasing the average $v \sin i$ measurement 
scales as $(v\sin i)_a^{-1}$, whereas the decrease in $\rho$ due 
to the over-estimate of $R_m$
is independent of $v\sin i$. The net effect is that the 
presence of unresolved binaries in a sample of slowly rotating
stars is to increase the inferred $\rho$ above its true value, but
binarity in a sample
of faster rotating stars will decrease the inferred $\rho$.

The Praesepe targets with measured $r\sin i$ appear to have relatively high
binary fractions (47 and 38 per cent in the upper and lower bins
respectively, see Section 4.2), but they are also fast rotators, with a
weighted mean $v\sin i \simeq 20$\,km\,s$^{-1}$.  Consequently binarity has
only a small net effect, leading to an {\it over-estimate} of $\rho$ by $\sim$1 per cent.

\nocite{Corsaro2017a}
\nocite{Jackson2013a}
\nocite{Kovacs2018a}
\nocite{Morales2009a}
\nocite{Torres2013a}
\nocite{Feiden2014a}
\nocite{MacDonald2013a}
\nocite{Jackson2014a}
\nocite{Covey2016a}
\nocite{Somers2014a}
\nocite{Hartman2010a}
\nocite{Jackson2013a}
\nocite{Lanzafame2017a}
\nocite{Baraffe2015a}
\nocite{Carpenter2001a}
\nocite{Douglas2017a}
\nocite{Jackson2009a}
\nocite{Horne1986a}
\nocite{Bagnulo2003a}
\nocite{Claret1995a}
\nocite{Skrutskie2006a}
\nocite{Barnes2007a}
\nocite{vanLeeuwen2009a}
\nocite{vanLeeuwen2017a}
\nocite{Feiden2014a}
\nocite{Feiden2015a}
\nocite{Somers2015a}
\nocite{Dotter2008a}
\nocite{Douglas2017a}
\nocite{LopezMorales2005a}
\nocite{Feiden2016a}
\nocite{Jeffries2017a}
\nocite{Somers2017a}
\nocite{Morales2008a}
\nocite{Mullan2001a}
\nocite{Kraus2016a}
\nocite{Bershady2008a}
\nocite{Jackson2014b}
\nocite{Cummings2017a}
\nocite{Bouvier2018a}
\nocite{Rhode2001a}
\nocite{Raghavan2010a}
\nocite{Duchene2013a}
\nocite{Yang2015a}
\nocite{Boesgaard2013a}
\nocite{Gaia2018a}
\nocite{Gaia2016a}
\nocite{Kesseli2018a}
\nocite{Wright2011a}
\nocite{Jeffries2011a}
\nocite{Newton2017a}
\nocite{Reiners2012a}
\nocite{Fang2016a}
\nocite{Gaia2018a}
\nocite{Jackson2010a}
\nocite{Jeffries2007b}
\nocite{Jackson2016a}
\nocite{Jackson2018a}
\nocite{boyajian2012a}
\nocite{Baraffe1998a}
\nocite{Spruit1986a}
\nocite{Kraus2017a}
\nocite{Kesseli2018a}
\nocite{Feiden2012a}
\nocite{Kraus2007a}
\nocite{Somers2015b}
\nocite{Wright2018a}
\nocite{Matt2015a}
\nocite{Garraffo2016a}
\nocite{Shulyak2017a}
\nocite{Chandrasekhar1933a}


\bsp 
\label{lastpage}
\end{document}